\documentclass{article}
\usepackage[utf8]{inputenc}
\usepackage{mathtools}
\usepackage{algorithm}
\usepackage[noend]{algpseudocode}
\makeatletter
\def\BState{\State\hskip-\ALG@thistlm}
\makeatother

\title{Spam Prevention Using zk-SNARKs for Anonymous Peer-to-Peer Content Sharing Systems}
\author{Alberto Inselvini}
\date{November 18, 2019}
\begin{document}
\maketitle
\begin{abstract}
    Decentralized unpermissioned peer-to-peer networks are inherently \newline
    vulnerable to spam when they allow arbitrary participants to submit content 
    to a common public index or registry; preventing this is difficult
    due to the absence of a central arbitrator who can act as a gate-keeper.
    For this reason indexing of new content even in otherwise decentralized 
    networks (e.g. Bittorrent with DHT, IPFS) has generally been left to 
    centralized services such as torrent sites.
    
    Decentralized methods for spam prevention, such as Web of Trust, already exist 
    \cite{Wernberg2016ScalableMF}\cite{883720}
    but they require submitters to assume pseudonymous identities and 
    establish trust over time. 

    In this paper we present a method of spam prevention that works under the assumption
    that the participants are fully anonymous and do not want different submissions of 
    theirs to be linked to each other.

    By \textit{spam} we do not specifically mean unsolicited advertising; rather it is the practice of 
    adding a large amount of content in a short time, saturating the capacity of the network,
    and causing denial of service to peers. The purpose of our solution is to prevent users 
    from saturating the system, and can be described as \textit{rate-limiting}.
    The system should be censorship resistant: it should not be possible for submissions 
    to be excluded because of the content itself, or for users to be excluded based on what
    they submit.

    We first discuss a solution based on a single, centralized rate-limiter that is
    censorship resistant and anonymous, then we extend this to a fully decentralized 
    blockchain-based system and present methods to make it economical and scalable.

\end{abstract}
\pagebreak
\section{Introduction}
    IPFS and Bittorrent provide decentralized protocols to share files and information, 
    addressed by their hash, without any need for centralization thanks to their DHTs.
    They are resistant to censorship and denial of service attacks.
    The caveat is that in order to retrieve a file, one must know its hash, which needs
    to be shared by users in some way that's external to the network, and generally centralized.
    \paragraph{}

    An example of this are torrent indexing sites, which provide search engines for finding
    the right torrents, and provide \textit{magnet links}, which are hashes of .torrent
    files. Torrent clients can use these hashes to obtain the files, which ultimately let 
    them retrieve the actual content the user requested.

    This model presents some problems: while the peer-to-peer network is unpermissioned,
    indexes are not; the owners of the site can decide which submissions will
    be included, and the sites themselves represent single points of failure.
    \paragraph{}

    Items retrieved through IPFS need not be files; they could be \textit{IPLD objects}
    representing, for example, posts on a bulletin board forum, in which case the use of a 
    centralized index represents a real opportunity for censorship.
    But it \textit{is} necessary for these posts to be added to a common feed in real-time
    for such an application to actually work.

    In fact, it's specifically this application that the present paper was intended for:
    an anonymous, unpermissioned and decentralized textboard-like forum.

    \paragraph{}
    The situation is such: we have a public append-only database that anyone can add entries to, 
    anonymously, with each entry being represented by its hash, which can then be used by peers to retrieve
    its content. All participants are peers and no one has special permissions such as the ability
    to remove entries.

    \paragraph{}
    Without an adequate system in place to prevent it, nothing stops a malicious peer from submitting a
    large amount of content in a short time, which can result in network congestion, as well as the 
    other peers' memory being filled with spam posts, burying legitimate content as well as overloading
    honest participants, whose resources are always limited.
    
    Because the network is both decentralized and anonymous it's not possible to reliably identify \textit{who}
    is causing this disruption, or distinguish genuine posts from spam automatically.

    Since nodes will relay messages for other nodes one cannot assume that messages coming from a particular
    IP address are being created by the peer running on that address.
    Additionally, anonymity networks such as Tor may be used by participants, in which case distinguishing
    different peers from each other may be impossible.
    
    Likewise, captchas cannot be used as they rely on centralized entities for verification.
    \paragraph{}
    Existing solutions \cite{Wernberg2016ScalableMF}\cite{883720} to this problem rely on users 
    maintaining pseudonymous identities through digital signatures and establishing them over time,
    if they are caught spamming, they can be "banned" by individual nodes.
    Such a solution obviously cannot work if the premise is that posts have to be anonymous,
    there should be no way to link submissions made by the same author to each other.

    We need a way to limit the rate at which an individual user can post (and possibly the total rate as well),
    without linking them to any of their posts.

\section{Centralized Rate-Limiter}
    Our first solution is based on a centralized and partially trusted rate-limiter,
    which will approve new entries to be added to the database.

    The rate-limiter will be trusted not to actively attack the network by spamming it,
    however we do not want to trust it not to enact censorship.

    For this reason it should not know what it is approving, nor which of its users have
    posted what content in the past.
    \paragraph{}
    The rate-limiter will decide the exact criteria it uses to limit posting.
    Users may be required to complete a \textit{captcha challenge} by the limiter in order to post, or it could
    simply impose its limits based on the IP address of the requester.
    The specific method used is not relevant to this discussion.
    \subsection{Naive Implementation}
        A naive solution would be to simply have the rate limiter sign the hash of an entry.
        This would make it impossible for the rate-limiter to discriminate based on content.

        The user would then submit his entry to the database along with the signature, which
        could be verified by every other peer.

        The problem with this approach is that the rate-limiter can read all posts retroactively,
        and through their hash, know which user submitted which one.
        This is because in the initial step the user sent his hash to the rate-limiter.
        This could lead to the user being banned, and it compromises anonymity.

    \subsection{On Tor}
        An astute reader might observe that if a user truly wants to be anonymous, they should
        just use Tor.
        
        Using Tor would be a requirement for secure anonymity regardless of how 
        we implement rate-limiting.
        The user needs to submit their entry to the database at some point and that involves
        potentially revealing their IP address to someone who is recording all traffic and timing 
        the submission of entries (this scenario is, by the way, completely outside of the scope of
        this paper; users who are concerned about a very powerful adversary \textit{should} be using Tor).
        \paragraph{}
        However, Tor is not very accessible. Using IPFS through the in-browser node doesn't involve installing
        any custom software on one's computer, nor using a separate browser.
        A system that requires Tor for operation is not going to be very popular.

        But there's another reason why Tor is not an appropriate solution (by itself).
        As we are going to see later, the system we propose has some very useful properties 
        that generalize beyond the centralized case, while Tor doesn't.
    
    \subsection{Salted Hash Commitment}
        As we have seen previously, revealing the hash of the content to the rate-limiter
        compromises our anonymity and needs to be avoided.
        \paragraph{}
        Instead of sending the hash of our entry \(E\), we generate a secret \(S\) and
        combine them into a \textbf{\textit{Merkle tree}} with two nodes, \(E\) and \(S\): 
        \[C := H(H(E) || H(S))\]
        Where \(H(x)\) is a hashing function, \(||\) indicates concatenation 
        and \(C\) is the root of the resulting Merkle tree, which we are going to call \textit{\textbf{Salted Hash Commitment}} (SHC).

        \textit{Salted}, because it is analogous to password hash salting, with the "salt" being represented by \(H(S)\).

        \textit{Commitment}, because it represents a \textit{\textbf{cryptographic commitment}} to \(E\). What this means is that
        \(C\) can only be generated from a particular \(E\), assuming that \(H(x)\) is secure. However it's not possible to derive \(C\) given \(E\),
        without knowing the secret \(S\).
        \paragraph{}
        This SHC is the value that we are going to reveal to the rate-limiter in order for it to be signed.

        Thanks to its properties, the problem with the previous "naive" implementation can be solved:
        the rate-limiter will sign \(C\) knowing that it can only be used to prove that a specific \(E\) was
        validated.
        \paragraph{Merkle tree} It may not be obvious to the reader why we are representing a salted hash as a Merkle tree.
        Thinking of this structure as a tree makes it easier to reason about making it part of a larger tree,
        which is something we are going to do later. Treating everything as a large tree will make 
        writing verification logic simpler, and it works well with blockchains, which use these data structures.

        \paragraph{Pedersen commmitments} Pedersen commitments are a type of cryptographic commitment more
        commonly used for this purpose,
        because they are additively homomorphic and work with low entropy messages.
        These properties are not needed in our case, which is why we opted for a generic cryptographic hashing
        function, which can be faster, and is more easily available.

    \subsection{Proof of Signing}
        After generating our commitment \(C\) and getting it signed by the rate-limiter with a signature \(\sigma\),
        we need to prove to all peers that our entry \(E\) was approved without compromising our privacy, 
        which means without revealing \(C\) or \(\sigma\).

        If we revealed either of these values, it would be easy for the limiter to connect our identity to
        \(E\) just like in the naive implementation, since they previously saw \(C\) when they signed it,
        creating \(\sigma\).

        \begin{algorithm}
            \caption{We define a function that, given arguments \(H(E)\), \(H(S)\), \(C\), \(\sigma\),
            and the limiter's public key \(K\) returns \(true\) if \(H(E)\) was approved, \(false\) if not.}\label{euclid}
            \begin{algorithmic}[1]
            \Procedure{VerifyApproval}{}
            \State $h_E \gets \text{hash of }E$
            \State $K \gets \text{the limiter's public key}$
            \BState \emph{private}:
            \State $h_S \gets \text{hash of }S$
            \State $C \gets \text{Salted Hash Commitment to }E$
            \State $\sigma \gets \text{signature of }C$
            \BState \emph{begin}:
            \If {$H(h_E||h_S) == C $ \textbf{and} $\textit{SignatureValid}(C, \sigma, K)$} 
            \State \Return $true$
            \EndIf
            \State \textbf{else} \Return $false$
            \EndProcedure
            \end{algorithmic}
        \end{algorithm}

        The problem we need to solve can be reduced to proving 
        to other peers that \(VerifyApproval(H(E), K, H(S), C, \sigma) \) 
        returns \( true \), without revealing the value of any of the arguments
        except for \(H(E), K\).
    
    \subsection{zk-SNARKs}
        \textit{Zero knowledge proofs} are methods by which one party can prove to another that 
        a statement about a value is true, without revealing that value.

        \paragraph{zk-SNARKs} (\textit{zero knowledge Succinct Non-interactive ARguments for Knowledge}) \cite{10.1007/978-3-642-17373-8_20} 
        are a family of zero knowledge algorithms that allow us to produce a proof that a given function
        returns a given value for a certain set of arguments, while only revealing some
        of these arguments and keeping the others private.
        \paragraph{}
        By applying zk-SNARKs to VerifyApproval() with our parameters we can generate the proof we need, 
        while only revealing \(H(E), K\).

        This proof can then be propagated to other peers, who will be able to verify it independently.
        By verifying it they will know that at some point \(E\) was approved by the owner of the key \(K\).

        They will not know anything else, as they won't be able to see any of the other arguments.
        This makes it impossible to directly connect any particular signing event to the entry itself, \(E\),
        or vice versa.

    \subsection{ZoKrates}
        In order for a function to be used with zk-SNARKs, it needs to be expressed in quadratic arithmetic
        form, which is not trivial to do.

        There are software packages such as \textit{ZoKrates} and \textit{snarky} which offer DSLs to 
        write algorithms in, as well as tooling that can automatically convert this code to the right format and
        generate proving and verifying programs for them.

        \paragraph{}
        Through the ZoKrates compiler, functions are turned into \textit{\textbf{arithmetic circuits}},
        which can then be used for zk-SNARKs once in quadratic arithmetic form.
    \subsection{Trusted Setup}
        After we compile our function into a circuit, we need to generate a pair of 
        public keys, the \textit{\textbf{proving key}} \(P_k\) 
        and the \textit{\textbf{verifying key}} \(V_k\).

        The former is used to produce proofs, the latter is used to validate them.
        \paragraph{}
        These keys, which are created only once, and are specific to the circuit they were made for,
        need to be used by every prover and verifier.
        They are generated in the so called \textit{trusted setup phase}.

        \paragraph{}
        In this phase we generate a random value \(\lambda\), then obtain \(P_k\) and \(V_k\) from a computation
        with \(\lambda\) and the circuit.

        \(\lambda\) is "cryptographic toxic waste": it should be eliminated immediately after this
        setup because it can be used to produce fake proofs for our circuit or reveal the private
        encrypted values in all proofs generated with \(P_k\).

        \paragraph{}
        The necessity of this procedure has generated controversy with some practical applications of
        zk-SNARKs, such as ZCash, a cryptocurrency that uses zero knowledge proofs to enable anonymous
        transactions.

        There is no way for a user of ZCash to trustlessly verify that the toxic parameters have been
        destroyed, and as such ZCash is not entirely trustless.

        However this doesn't imply that the anonymity of transactions
        could be violated, even if one knew \(\lambda\).

        zk-SNARKs have three fundamental properties \cite{inproceedings}, the latter two of which 
        are contingent upon the attacker not knowing the secret parameter:
        \paragraph{Completeness} Given a statement and a witness, the prover can convince the verifier.
        \paragraph{Soundness} A malicious prover cannot convince the verifier of a false statement.
        \paragraph{Zero-knowledge} The proof does not reveal anything but the truth of the statement,
        in particular it does not reveal the prover’s witness.
        \paragraph{}
        It's impossible to create a SNARK algorithm where, with knowledge of \(\lambda\), \textbf{both}
        \textit{Soundness} \textbf{and} \textit{Zero-knowledge} are preserved \cite{cryptoeprint:2016:372};
        however an algorithm can be resistant to \textit{subversion} for soundness (\textbf{S-SND}) 
        \textbf{or} zero-knowledge (\textbf{S-ZK}).

        ZCash uses \textit{Groth16} which is S-ZK. That means that while an attacker could potentially create
        fake proofs, generating ZCash cryptocurrency out of thin air, they could never reveal private information,
        such as transaction amounts or the parties involved.

        \paragraph{}
        By using an S-ZK SNARK, privacy is not contingent on the assumption of a trusted setup:
        in our case the only thing that could be done by the rate-limiter if it generated the keys and
        didn't destroy the parameter is generating fake proofs for its own approvals, which is obviously
        pointless.

    \subsection{Prepared Flood Attack}
        Our Zero Knowledge scheme allows us to separate the approval of an entry from the entry itself,
        while still being able to arbitrarily limit how much a user can post in a given time period.
        
        However this leaves the system open for a rather dangerous attack: a malicious peer could 
        prepare a large amount of approvals over a long period of time, by requesting them from the
        rate limiter at the normal allowed rate, accumulating them, then releasing them all at once.
        
        This attack is hard to detect and it's impossible to tell which messages were approved long ago
        and which messages are recent and honest.

        \paragraph{}
        In order to prevent this, we need to add a time element to our rate-limiting.

    \subsection{Timestamped Approval}
        We are going to change the approval process by adding a new step.

        After receiving the Salted Hash Commitment from the user, the limiter
        creates a larger Merkle tree which includes the SHC and the current timestamp,
        signs its root, and then returns the tree and signature to the user.

        \[ A := H(C||H(T)) \]

        Where A is the new Merkle root, which we will call Timestamped Approval,
        \(C\) is the SHC, and \(T\) is the current timestamp.
        \paragraph{}
        After signing \(A\) we need to send it and its signature to the user along with \(T\),
        which will be necessary for our new updated proof scheme.
    
    \subsection{Adding Time to Our Proof}
        We need to construct a different proving circuit while still maintaining the disassociation
        between approval and entry necessary for our privacy.

        For this reason it's important that we don't disclose the exact timestamp \(T\) we received from the rate-limiter.
        Instead we will add a new public parameter to the proof, timestamp \(T' != T\), which can differ from \(T\) 
        by \textit{at most} a small amount \(dT\), such that
        \[T - T'= \pm dT\]

        \begin{algorithm}
            \caption{We define the new version of VerifyApproval}
            \begin{algorithmic}[1]
            \Procedure{VerifyApproval}{}
            \State $h_E \gets \text{hash of }E$
            \State $K \gets \text{the limiter's public key}$
            \State $T' \gets \text{public timestamp}$
            \BState \emph{private}:
            \State $T \gets \text{private timestamp from rate-limiter}$
            \State $h_S \gets \text{hash of }S$
            \State $C \gets \text{Salted Hash Commitment to }E$
            \State $A \gets \text{Timestamped Approval for }A$
            \State $\sigma \gets \text{signature of }A$
            \BState \emph{begin}:

            \If {$H(h_E||h_S) == C $ \textbf{and} $H(C||H(T)) == A$ \textbf{and} $|T-T'| <= dT$ \textbf{and}  $\textit{SignatureValid}(A, \sigma, K)$} 
            \State \Return $true$
            \EndIf
            \State \textbf{else} \Return $false$
            \EndProcedure
            \end{algorithmic}
        \end{algorithm}

        This allows us to show a public timestamp that differs from \(T\), without letting any 
        malicious user lie about \textit{approximately} when \(E\) was approved.

        Posters should simply set \(T'\) to a random value that falls within \(T \pm dT\).
    \subsection{Choosing \(dT\)}
        In choosing the value of \(dT\) we have to strike a balance between privacy and security against attacks.
        \paragraph{}
        If the value is too large, then we have the same problem we had without timestamps, where we don't know
        when an Entry was approved for inclusion.

        If it is too small, then we risk compromising the user's anonymity: if a post was approved at a time \(T\),
        and no other entries are approved between \(T-dT\) and \(T+dT\), then once the content is published
        with public timestamp \(T'\), it's completely obvious to the rate limiter which user made the post,
        since it can only be one.
        \paragraph{}
        However, if at least two peers post in this timeframe \(dT\) it's impossible to know which one of them 
        published what.
        \paragraph{}
        The choice should be made while taking into consideration the nature of the application, as well as
        the number of users that will be contributing at any given time.

        In any case, if our rate limiting rule is that users are allowed to make one contribution for every
        time period of \(T_l\) seconds, setting \(dT := T_l\) should always be safe against flood attacks;
        there is no reason for it to be lower than that.

    \subsection{Defeating Attackers}
        While we've successfully managed to limit how much an attacker can \textit{spoof} their
        timestamp, we haven't actually described a method to use this information to prevent attacks.

        How exactly this is accomplished will depend on the nature of the database and the data stored within.
        
        If it's a "feed" type of store, such as the bulletin board forum described in the Introduction,
        then old entries may just be ignored, as they are irrelevant. Posts past a certain time would 
        eventually be deleted in any case.

        Otherwise, clients may simply put all incoming data in a queue, prioritizing the most
        recent timestamps and putting the least recent in the back of the queue.

        If the queue becomes too large, the oldest items are dropped.

        By using this method we ensure that recent items, posted by non-malicious users, are processed first,
        and the attack doesn't compromise the regular functioning of the network.
    
    \subsection{Conditions for Effective Anonymity}
        With this system, we can guarantee that it's not possible for the rate limiter to tell with certainty which posts
        belong to which users, as long as different users are posting at the same time; the \textit{anonymity set} 
        for a peer is constituted by the peers who posted in the same \( \pm dT \) timeframe.

        Assuming a consistent stream of submissions, users could even authenticate themselves to the rate-limiter 
        in order to post, with a more persistent identity than their IP address, and still be anonymous.

        This opens up the possibility for a community with restricted membership where individual contributions are
        not linked to any particular account.

        \paragraph{}
        While it's not possible to link entries their authors \textit{with certainty} under these conditions,
        in a real-world scenario, total anonymity would be all but guaranteed; for example, the content itself
        could link different posts together, it could even hold information sufficient to obtain the author's
        real-life identity.

        The FBI has identified criminals solely from them discussing current local weather conditions
        on public online chats.
        \paragraph{}
        The primary purpose of these methods is to make it unfeasible for the rate-limiter to act as a moderator,
        rather than guarantee the safety of our users.

        The secondary purpose is to allow, for example, anonymous discussion between people, not necessarily to \textit{hide}
        anything, but because the author believes that such conditions foster a better environment and allow people to be
        more honest with each other.

        For this reason it is sufficient that de-anonymization be impossible to do automatically, and difficult to do manually.
        Resistance against any type of targeted attack is out of scope.
    
    \subsection{Fungible Approvals}
        Until now we have always made sure that our approvals are not \textit{fungible}, i.e. interchangeable with each other.
        Each one is only valid for the specific entry it was made for, as it is based on a commitment to that entry.
        \paragraph{}
        But we could do this differently, and allow the approvals to be fungible.

        Everything would work the same, except that in place of the entry's hash \(H(E)\), we use a unique random number \(Q\), 
        the \textit{nonce}, with enough entropy such that no other peer might generate the same one by accident.

        \(Q\), just like \(H(E)\) before, is used in the generation of the "commitment", is hidden from the validator,
        and later constitutes one of the public arguments to the proof, visible to every node that gets our entry.

        The difference is that previously other than checking if the proof was valid, nodes would ensure that the hash argument
        it coincided with the hash of the entry itself.

        Now, instead, they need to check if they've seen \(Q\) before, to make sure that the same proof isn't used twice.
        This is why Q needs to be unique. If they received another message with the same nonce previously, they reject the new message.

        We call this system \textit{token based}, because it is as if the user received a fungible token from the verifier,
        spendable by posting.

        \paragraph{}
        A token system has one major advantage in terms of privacy: previously we could only get approvals from
        the rate limiters \textit{after} we decided what content to submit, in the case of a forum that would mean
        \textit{after writing a post}.

        Tokens can be obtained in advance, especially with a more generous \(dT\) value.
        If, for example, \(dT\) was one day long, then we could get our tokens at the beginning of the day,
        and spend them in posting throughout the next 24 hours.

        The anonymity set would be much greater since our posts could belong to anyone who has requested tokens
        in the previous 24 hour period.
        \paragraph{}
        The issue with this model is that it creates a \textit{double spending} problem.
        If a user decides to use the same token twice, for two different entries, peers will decide which 
        of his posts to reject based on which one they receive first.

        This means that if two nodes receive the two posts in a different order, they will reject a different one.
        This leads to inconsistent state between nodes, which might or might not be a problem depending on the application.
        \paragraph{}
        Inconsistent state can be avoided by using a common metric to decide which message is considered valid.
        For example, comparing the hashes and choosing the greater.
        Or, discarding all messages with the same token in case of a double-spend, including the original.
        
        The compromise here would be in the \textit{finality} of our messages: the former option allows a user to effectively
        replace something they previously sent, the latter allows them to delete it entirely.
        \paragraph{}
        The double spending problem can be solved by using a blockchain in place of a centralized rate-limiter, however.

        Another issue is that a long grace period \(dT\) can create problems with some kinds of applications, besides
        the potential for flooding.
        For example, in a forum we don't want to allow users to create posts in the past, altering the history
         of discussions.
        This can also be solved with a blockchain.

        \paragraph{Front-Running} The somewhat naive, but easy to understand implementation of a token based system 
        that we have illustrated in this section can allow attackers to \textit{front-run} honest users.

        Proofs and their corresponding "tokens" are no longer tied to any particular entry hash.
        So the malicious user might receive a post from an honest user, \textit{steal} their proof, and use it for
        their own content, sending it immediately to as many peers as possible.

        To these peers this just looks like a double spend. Depending on what they see first, they might end up rejecting
        the honest user's entry.

        This issue can be rather trivially solved by including the entry's hash as a public parameter during the generation
        of the proof. This hash would still not be present in the commitment, so it's not necessary for it to be known
        before getting approval from the rate-limiter. The "token" remains fungible, but the proof itself isn't.

        Doing this ensures that the proof is only valid for a particular post.
        \paragraph{}
        Alternatively, the nonce could be the hash of a public key generated on the fly by our user, different for each token.
        We add the requirement that each entry must also be signed by the private key corresponding to the nonce.

        This signature ensures that a different message cannot use our token, unless the attacker has our private key.
        Since each message uses a unique key pair, there is no loss in privacy.

        The main advantage of using signatures is that the proofs can be generated before the messages are written. As this is a
        compute intensive process, doing it ahead of time might constitute a significant optimization in user experience.

        \pagebreak

\section{Federated Model}
    The model discussed in the previous section involves a centralized, trusted arbiter which is a problem for a
    fully decentralized system.

    The arbiter could act maliciously by spamming the network, or cease to function completely.

    If this happened, individual nodes could decide to trust a new arbiter, but that requires
    everyone to make the switch at the same time, or the common database will end up being inconsistent between peers.

    \paragraph{}
    In this section we discuss how the previous ideas can be extended to a federated model with several rate-limiters.
    \subsection{Trusted Federation}
        In a Trusted Federation, all nodes choose to trust a certain set of rate-limiters, and accept entries approved by any of
        their signatures.

        This system only adds redundancy to the one-arbiter model, and it requires everyone to have the same set of accepted keys,
        otherwise there will be a split in the network where certain peers don't accept some entries because they were
        approved by an untrusted rate-limiter.
        \paragraph{}
        Additionally, the various validators need to trust each other and share information about their users in order to
        rate limit effectively.

        If they act independently, a malicious poster could rotate between them, asking each to approve a different entry,
        leading to him being able to post \(n\) times as many entries in a span of time compared to what he should
        be allowed to post, where \(n\) is the number of validators.
    \subsection{Web Of Trust Federation}
        An alternative to a fully trusted federation is one where inclusion in the set of validators is contingent on being
        part of a so-called \textit{Web of Trust} \cite{883720} (or any similar scheme \cite{Wernberg2016ScalableMF}), where one's identity can be established over time and gradually 
        receive trust from the rest of the network. Acting maliciously causes one to be removed from this network.
        \paragraph{}
        As the reader might recall, we previously mentioned this model in the context of it being applied to 
        individual users, but rejected it on the basis of our requirement for anonymity.

        This requirement does not exist for the rate-limiters, so such a system is an option.

        One could see this as analogous to a decentralized network based on \newline 
        pseudonymous identities, but where these validators essentially act as \textit{proxies} for the users to post anonymously.
        It would be as if the posts were \textit{made} by the rate-limiters approving them.

        If one of them approves too many posts in a given amount of time, they are removed from the Web Of Trust.

        \paragraph{}
        There are a number of issues with this approach.
        \paragraph{}
        The removal of a validator needs to happen deterministically for all users involved or, as explained above, the
        network splits. In that case, a node might inadvertently use a removed validator when posting, leading to his posts
        being ignored by peers.
        
        This is difficult to accomplish if removal is based on the number of posts approved in a given time, since a peer could
        be offline when the attack happens.
        \paragraph{}
        Another problem is that we cannot know \textit{how many different users} are behind the entries approved by a certain 
        validator.
        By setting a hard limit, a validator which receives many legitimate requests may be forced to make them wait
        for a long time before approving. A possible solution is to base this limit on how much the validator is trusted:
        a validator that has been around for longer can approve more posts; the problem is that the exact number
        needs to be known to them, and it needs to be agreed upon by all users.
        \paragraph{}
        Lastly, as explained in the previous section, if the validators don't share information about users this can be exploited
        by a malicious user wishing to post far more than they are allowed.

        However, if each rate-limiter has a limit for how many total posts they can approve in a given timeframe, this problem is avoided.

    \subsection{Limitations of the Federated Model}
        As we have seen, it is difficult to operate a federated network while maintaining the kinds of guarantees 
        that one would expect from a decentralized network; however this model is still in some ways better than 
        other federated protocols, such as ActivityPub (mainly known for its most popular implementation, Mastodon).

        For example, it's not possible for members of the federation to censor users, and no one is locked to 
        any particular validator because posts are anonymous.

        On Mastodon, moving to a different \textit{instance} involves creating a new account on the instance one is moving to,
        which is separate from the previous. This causes a lot of friction when transferring.

\section{Blockchain Model}
    In order to make our rate-limiting fully decentralized, we can replace our trusted validators with a blockchain, such as
    Ethereum.

    Instead of getting our commitments signed we create a transaction that adds the SHC to the blockchain's state.
    Now our zero knowledge proof won't verify any signatures, and will instead prove that there is a Salted Hash Commitment to our entry
    included in the current block's state Merkle tree.
    
    Since the SHC is itself a Merkle tree, this can be reduced to proving that our post hash is now part of the last block's tree.

    Since every block is timestamped, there is no need to include a timestamp like before.

    \paragraph{}
    A blockchain limits the amount of posts a user can make because adding a value to the blockchain's state is expensive.
    Also, there are only so many transactions that can fit into a block, which limits how much can be posted in a given time period.

    The cost of a transaction is too high for many applications of this rate-limiter however, and its throughput
    might pose too much of a limit to how fast entries can be added to our common database.

    We will show how to implement blockchain-based rate limiting in greater detail in the following paragraphs, and present
    a solution to the economical and throughput issues.
    \subsection{Smart Contract}
        The Ethereum blockchain is used in our example.
        We create a trivial smart contract with two methods: 
        
        A unary function, \(append(h)\), which takes a string representing a hash as its only argument,
        and appends it to an array, \(D\), the only state the smart contract keeps internally.
        It may optionally fire an event whenever a new value is added, in order for clients to be able to track
        when this happened.

        A nullary function, \(get()\), which takes no parameters and returns the current value of \(D\).

        \paragraph{}
        We can now use a transaction to add a hash to \(D\), stored on the blockchain.

        Any peer can query the smart contract and get the latest value of \(D\), in order to check if a hash
        is included in it.

        They can also listen to events and update their copy of \(D\) accordingly, removing the need to query the contract
        every block.
    \subsection{Collectors}
        In order to avoid requiring one transaction to be made for each entry added to our database, we introduce the collector.

        A collector takes the place of the rate-limiter from our previous models.

        Collectors receive Salted Hash Commitments from users, combine them all together in a larger Merkle tree,
        and add its root \(R\) to the blockchain by calling \(append(R)\).

        They then return to each user the Merkle proof of inclusion in \(R\) of their SHC.

        \paragraph{}
        Users can now produce a zero knowledge proof that their entry \(E\) is included in the larger tree with
        root \(R\).

        This proof would have \(H(E)\) and \(R\) as public parameters, \(H(S)\) and the Merkle proof of inclusion in \(R\)
        as private parameters.

        Peers will consider an entry approved if the proof is correct, the hash corresponds to the entry's hash,
        and \(R\) is in \(D\).
    
    \subsection{Collector Economics}
        Through the collector system many different entries can be approved with a single transaction, while storing a
        single hash on the blockchain.

        Although it still costs money to approve posts, and a transaction has to happen any time one (or more) posts are
        approved, the \textit{marginal cost} of each individual entry is lowered to a fraction of a transaction's.
        The total \textit{throughput} in terms of posts per block that the blockchain can support is also no longer
        limited to any particular value, because the blockchain space required is constant for any number of entries.

        \paragraph{}
        This constitutes a double edged sword; we don't actually want the number of entries added per block to be unlimited,
        and we do not want to allow any malicious user to add a theoretically infinite amount of content while only 
        paying a constant amount.
        \paragraph{}
        In order to solve these issues we need to limit the amount of entries that can be added per transaction,
        and we do so by limiting the \textit{depth} of the Merkle tree \(R\).

        The total number of entries that can be stored in a Merkle tree is equal to \(2^n\), where \(n\) is its depth.
        As an example, we limit the tree's depth to 3, which means 8 entries approved for each transaction on the blockchain.

        In order to accomplish this we write our proving function such that it only works with a tree of depth 3.

    \subsection{Blockchain Privacy}
        While a public blockchain such as Ethereum is permissionless, and as such satisfies our requirement to make
         censorship impossible, it raises the issue of privacy.
         \paragraph{}
        Thanks to collectors, individual users do not have to make transactions on the blockchain themselves,
        which reduces this concern. 
        
        Collectors cannot know for sure what commitments correspond to which posts,
        however the timing can give them clues if there are a low amount of submissions over a certain period. 
        In practice, the anonymity set might be small.
        \paragraph{}
        Therefore it is important for users to keep their identities hidden from the collector when they make a submission.
        However, collectors might require small payments to compensate them for the transaction fees they pay. 
        
        Because of this, the availability of a payment system that preserves the privacy of the participants
        (unlike most current public blockchains, where all transactions are public) is necessary for our scheme to function as intended.
        \paragraph{}
        It should be noted that collectors can accept payments from other blockchains that have privacy features, 
        but don't support smart contracts, like Monero (XMR) or ZCash (ZEC).
        The only issue is that this doesn't protect the privacy of the collector itself.

\section{Conclusion}
        In this paper, we have discussed various approaches for rate-limiting anonymous submissions of information to a
        decentralized, permissionless system, with the objective of preserving these three properties of anonymity, decentralization, 
        and permissionless access, while still retaining effectiveness at preventing attackers from flooding our network.

        We have made use of zero-knowledge proofs to accomplish this goal, and have discussed the benefits and drawbacks to various
        implementations of our scheme, such as a centralized approach, a federated model, and a fully decentralized system that leverages
        modern blockchains.

        We have shown that while it is theoretically possible for a centralized or federated system to meet the requirements of 
        being permissionless and anonymous, both of these models present issues that can be avoided with the use of a blockchain ledger.

        The ability to execute private transactions on a smart-contract-capable public blockchain is currently the most important missing piece 
        of technology for our solution.

    \pagebreak

\bibliography{citations}
\bibliographystyle{ieeetr}
\end{document}